# Between Myth and History: von Neumann on Consciousness in Quantum Mechanics

Federico Laudisa

Department of Humanities, University of Trento

Via T. Gar 14, 38122 Trento, Italy

**Abstract**

The von Neumann attitude on such a deep interpretational question as the role of a human observer in order for the quantum description of measurement to be consistent has been long misrepresented. The large majority of the subsequent literature ascribed to von Neumann a radical view, according to which not only the collapse was in itself a truly physical process, but also the only way to accomodate it within a quantum description of a typical measurement was the introduction of human consciousness as a kind of 'causal' factor. Inspired by the work of reconstruction pursued by the phenomenological reading of the London-Bauer approach, started by Steven French more than twenty years ago, the account I propose substantiates a significantly more cautious attitude by von Neumann: the time seems then ripe to tell a more balanced story on the relation between the notion of consciousness and the foundations of quantum mechanics in the work of the first scientist – János von Neumann – who explicitly and rigorously addressed the implication of a really *universal* formulation of quantum physics.

# 1 Introduction

Some twenty years ago Steven French undertook the task of re-interpreting a work that had been confined to one of the weirdest areas in the debates on the foundations of quantum mechanics. The work, published in 1939, was a sort of mini-textbook written by Fritz London and Edmond Bauer on the mathematical and conceptual foundations of quantum mechanics (London, Bauer 1939), whose fortune relies essentially on its being credited with a most peculiar claim: the idea that the notion of consciousness plays a major and active role in making sense of the quantum process of measurement. On the basis of some historical evidence concerning an influence of the Husserlian phenomenology on London's view of physics, French could propose a new, detailed reading of the London-Bauer approach, in which the role of the notion of consciousness turned out to be much more nuanced than just being a sort of 'mental trigger' for the so-called collapse of the quantum state to take place (French 2002, 2020, 2023). French aims at showing that, contrary to a sort of consensus until then, the London-Bauer approach does *not* support the claim that it is consciousness itself that *empirically causes* the collapse of the quantum state, since the notion of consciousness employed in their work should be interpreted along the lines of a Husserlian, phenomenological approach (in which such a 'causal' role of consciousness would make no sense). In the folklore view, widely shared until not long ago, the source of the idea that it is consciousness that brings about the collapse was traced back to von Neumann's 1932 treatise on the mathematical foundations of quantum mechanics (von Neumann 1955). Just to provide an instance, in a short historical section of their book *The Quantum Theory of Measurement* (1991), Paul Busch, Pekka Lahti and Peter Mittelstaedt wrote that

> [t]he London-Bauer (1939) theory of measurement and its critique through the story of Wigner's (1961) 'friend' are concerned with the possibility already pointed out by von Neumann and Pauli that the observer's consciousness enters in an essential way into the description of quantum measurements." (Busch, Lahti, Mittelstaedt 1991, 2-3).

Since in the above mentioned folklore view the fate of the London-Bauer work was strongly intertwined with the alleged von Neumann view, taking seriously the French reconstruction implies in turn the need to carefully reconsider the status of von Neumann's treatment of the quantum theory of measurement. In particular: is it *really* clear what the von Neumann stance was about the collapse of the quantum state as an *actual* physical process? In the present paper I would like to point out not only that the textual evidence does *not* allow to single out a clear formulation, that von Neumann might have defended, but also that the status of the collapse of the quantum state as an actual physical process is in a conceptual tension with a formal consequence of the overall framework that von Neumann developed for the analysis of measurement as a process totally governed by the laws of quantum mechanics. As a consequence, if it is far from clear what, according to von Neumann, the status of the collapse was supposed to be, even less clear was the view that the collapse is in fact 'caused' by consciousness. On the basis of these conclusions, a more cautious and nuanced reading of von Neumann's presuppositions concerning measurement in quantum mechanics needs to be developed, along lines that are similar to those pursued by the French phenomenological reading of the London-Bauer approach.

The plan of the paper is the following. In the section 2, on the background of the French interpretation of the London-Bauer approach, I would like to review some of the history of quantum physics literature concerned with the consciousness-collapse relationship in von Neumann's model of the quantum measurement process, in order to see the terms in which the thesis of a strong belief by von Neumann in the collapse as an actual physical process triggered by consciousness has been considered, in fact, an established view. In the section 3, I will focus on the von Neumann model itself, in order to support three points. *First*, von Neumann was far from having a well—developed argument in favour of the collapse as an actual physical process; *second*, a major result of his model of the quantum measurement process stands in real tension with a possible argument like that; *third*, as a consequence of the first two points, the attribution to von Neumann of the belief in a consciousness-induced collapse is far from being clear and uncontroversial. Finally, in support of the above points, in section 4 I will focus on two 'indirect' sources that may speak in favour of the analysis developed in the section 2 and 3. The first is the account of the quantum theory of measurement that David Bohm provides in his 'Copenhagenish' textbook (Bohm 1951). This account turns out to be extremely interesting in being a sophisticated version of the standard interpretation and – *at the same time* – in being totally resonant with the caution in von Neumann's account about the nature of the quantum state collapse and the role of consciousness in the measurement process. The second is the well-known Wigner 1961 article on what he calls *the mind-body question*, and in which he introduces the 'Wigner friend' argument. Being a classic reference on the apparently unavoidable role of consciousness for the consistency of the account of the measurement process in quantum mechanics, this article should have been the natural place for Wigner to include von Neumann among the supporters of such a view: that this is *not* the case may be a further element to corroborate the point that, as a matter of historical fact, von Neumann *did not* develop any theory of consciousness as a pillar of a possible quantum theory of measurement.

## 2   Von Neumann on Consciousness and the Collapse of the Quantum State: the Received View

The issue of whether quantum mechanics should be applicable to *all* physical systems, including measuring instruments, has been indicated as the problem of the *universality* of quantum mechanics. Universality for quantum mechanics – or 'quantum fundamentalism', as for instance Zinkernagel calls it – is the claim that "Everything in the universe (if not the universe as a whole) is fundamentally of a quantum nature and ultimately describable in quantum-mechanical terms." (Zinkernagel 2015, 419). Such problem in principle emerged since the very origins of quantum theory. As London and Bauer emphasized in the Preface to their 1939 text:

> The majority of introductions to quantum mechanics follow a rather dogmatic path from the moment that they reach the statistical interpretation of the theory. In general, they are content to show by more or less intuitive considerations how the actual measuring devices always introduces an element of indeterminism, as the interpretation demands. *However, care is rarely taken to verify explicitly that the formalism of the theory, applied to that special process that constitutes the measurement, truly implies a transition of the system under study to a state of affairs less fully determined than before. A certain uneasiness arises*. One does not see with what right and up to what point one may, in spite of this loss of determinism, attribute to the system an appropriate state of its own. Physicists are to some extent sleepwalkers, who try to avoid such issues and are accustomed to concentrate on concrete problems. But it is exactly these questions of principle which nevertheless interest nonphysicists and all who wish to understand what modern physics say about the analysis of the act of observation itself. (London, Bauer 1939, 218-219, emphasis added).

As a matter of fact, in the early days of the debate on the foundations of quantum mechanics it was far from clear what the relation between the classical and the quantum regimes in the event of a measurement was supposed to be: even two important and highly influential works – such as *Die Physikalischen Prinzipien der Quantentheorie* (1930) by Werner Heisenberg and *The Principles of Quantum Mechanics* by P.A.M. Dirac (1930) – although aspiring to provide a general and rigorous framework for quantum mechanics, did not include an explicit analysis of the measurement process in terms of the universality of quantum mechanics [1].

The London-Bauer text aims exactly at a novel contribution in this respect, but the authors refer shortly and vaguely to the von Neumann treatment in his 1932 treatise:

> Although these problems have already been the subject of deep discussions (*see especially von Neumann, 1932*), there does not yet exist a treatment both concise and simple. This gap we tried to fill. (London, Bauer 1939, 219, emphasis added).

The formal model of measurement that von Neumann provided in his 1932 treatise is actually an explicit 'implementation' of the thesis on the universal nature of quantum mechanics: in addressing the implications of the assumption that – in the context of a measurement of a physical quantity on a quantum system $S$ with an apparatus $A$ – the laws of QM govern *both $S$ and $A$*, the model in fact led physicists to realize how controversial the status of measurement in quantum mechanics would have been[2]. It is probably a paper by von Neumann on the statistical structure of the theory (von Neumann 1927), however, that represents the first analysis of the quantum measurement process in a direction that could pave the way to the hypothesis of the universality of quantum mechanics[3]. The 1927 results were included by von Neumann

---

[1] For a detailed analysis of the reasons that motivate the lack of awareness on a 'measurement *problem*' in the classic works of Heisenberg 1930 and Dirac 1930, see Laudisa 2024. A truly comprehensive historical reconstruction of the stages through which measurement in quantum mechanics has become a widely recognized problem is, in fact, still missing: for a recent attempt, albeit partial and very concise, see Pessoa Jr 2022.

[2] Within the project of providing a scientifically respectable picture of the classical-quantum relation involved in measurements, the occurrence of a sort of 'reduction' process under specified conditions had been mentioned already by Heisenberg in his 1927 uncertainty paper. In the section 3 of the paper (entitled "The transition from Micro- to Macromechanics"), Heisenberg describes a sequence of position measurements on electrons, focusing on the precision allowed by the use of light to determine position: at every step "the results of later measurements can only be calculated when one again ascribes to the electron a 'smaller' wavepacket of extension λ (wavelength of the light used in the observation). Thus every position determination *reduces the wave packet* back to its original extension λ." (Heisenberg 1927: 74, emphasis added).

[3] In a review paper published in 1958 in the *Bulletin of the American Mathematical Society*, the Belgian physicist Léon van Hove stressed the importance of that pioneering paper, but mentioned a connection with Niels Bohr that in fact is quite obscure: "In the same paper von Neumann also investigates a problem which is still now the subject of much discussion, viz., the theoretical description of the quantum-mechanical measuring process and of the noncausal elements which it involves. Mathematically speaking von Neumann's study of this delicate question is quite elegant. *It provides a*

in a wider framework in his 1932 treatise on the mathematical foundations of quantum mechanics[4], in which the assumption of existence within the theory of two distinct modalities of evolution over time for the states of quantum systems was explicitly introduced. The first is the unitary evolution (called "causal change", for instance, at page 417 of von Neumann 1955) governed by the time-dependent deterministic Schrödinger equation – holding between measurements – whereas the second is the non-unitary, irreversible, stochastic evolution induced by measurements. In Chapter III ("The quantum statistics") von Neumann describes the latter modality of evolution as follows. Let $R$ a self-adjoint operator representing a physical quantity ℛ, to be measured on a quantum system in the state $\psi$. Under standard mathematical constraints on $R$,

> a measurement of ℛ has the consequence of changing each state $\psi$ into one of the states $\phi_1, \phi_2,...$, which are connected with the respective results of measurements $\lambda_1, \lambda_2,...$ The probabilities of these changes are therefore equal to the measurement probabilities for $\lambda_1, \lambda_2,...$ (von Neumann 1955, 216).

If this is the *description* of the above mentioned non-unitary, irreversible, stochastic evolution induced by measurements, what is the *cause* or the *explanation* of such evolution?[5] Few lines later, von Neumann explicitly declares his agnosticism:

> We have then answered the question as to what happens in the measurement of a quantity ℛ, under the above assumptions for its operator $R$. *To be sure, the "how" remains unexplained for the present.* This discontinuous transition from $\psi$ into one of the states $\phi_1, \phi_2,...$, [...] is certainly not of the type described by the time dependent Schrödinger equation." (von Neumann 1955, 217, emphasis added).

---

*clear-cut formal framework for the numerous investigations which were needed to clarify physically the all-important implications of quantum phenomena for the nature of physical measurements, the most essential of which is Niels Bohr's concept of complementarity.* (van Hove 1958, p. 97, emphasis added).

[4] All quotes and page references are taken from the English edition (von Neumann 1955).

[5] In the 1927 paper von Neumann gives an early description of what was to become the collapse of the quantum state: "A measurement is in fact in principle an intervention (from which the 'acausal' character of quantum mechanics arises, cf. Heisenberg, Zeitschrift für Physik, Vol. 43, 3/4, 1927); however, we must assume that the alteration proceeds in the interest of the experiment, i.e. that *as soon as the experiment is carried out, the system is in a state in which the same measurement can be executed witout any further alteration of the system. Otherwise stated: if the same measurement is carried out twice (with nothing happening in the interim!), the same result is obtained.* (von Neumann 1927, 97, fn. 30, emphasis added).

How did this agnosticism turn mysteriously into a view in which the 'explanation' of the collapse relies on nothing less than the consciousness of a human observer?

In an article published in 1963 on the *American Journal of Physics* and devoted to a review of the main accounts of measurement in quantum theory at the time, Abner Shimony had expressed an ambivalent evaluation:

> The most systematic theory of observation in quantum mechanics was proposed by von Neumann and later presented more simply (and in some cases more deeply) by London and Bauer. They consider transitions of type 1, discontinuous transitions due to the performance of a measurement, to be an ineliminable aspect of quantum theory; *and they explicitly understand by 'measurement' the registration of the result in a consciousness*. (Shimony 1963, 757, in Shimony 1993, 7, emphasis added).

Putting Shimony's assimilation of von Neumann and London-Bauer aside for a moment, we have to recognize that this statement is still rather prudent, since the emphasized passage suggests for the consciousness simply the task of 'storing' the registration of an experimental result. Moreover, a couple of pages later, in a section significantly entitled *The Ability of the Mind to Reduce Superpositions*, Shimony writes explicitly that "von Neumann says *almost nothing* about the consciousness of the observer" (Shimony 1963, in Shimony 1993, 7, emphasis added). Unexpectedly, however, in the concluding section Shimony goes much further, ascribing to von Neumann (and London-Bauer) the more radical view according to which consciousness has a truly causal role:

> von Neumann, London and Bauer propose that objective changes of the state of physical systems occur when certain measurements are performed, and they explain these changes by referring to the subjective experience of an observer. This mutual involvement of physical and mental phenomena is counter-intuitive in the extreme, though without apparent inconsistency. However, there is no empirical evidence that the mind is endowed with the power, *which they attribute to it*, of reducing superpositions; and furthermore, there is no obvious way of explaining the agreement among different observers who independently observe physical systems. Thus, their interpretation of quantum mechanics rests upon *psychological presuppositions* which are almost certainly false. (Shimony 1963, 772, in Shimony 1993, 31, emphasis added).

It is quite clear that the statement, according to which von Neumann says *almost nothing* about the consciousness of the observer, can hardly be made consistent with the statement according to which the very same von Neumann is adopting such bold 'psychological presuppositions' so as to allow him to endow the mind of the power of triggering the collapse of quantum states! Apparently Henry Margenau had been the first to introduce the term *projection postulate* in a paper published on *Philosophy of Science* in 1958 [6], and the same Margenau had confirmed in 1963 von Neumann's uneasiness about the status of the collapse:

> Brief comment will here be made on a "postulate" introduced by von Neumann, recognized by him, I believe, as pictorially useful and consistent with the axioms of quantum mechanics (as of course all special kinds of interaction, whether or not they qualify as measurements, must be) but *also as avoidable*. (Margenau 1963, 476, emphasis added).

An inconsistency similar to that we pointed out in the Shimony 1963 article occurs in the Max Jammer 1974 book *The Philosophy of Quantum Mechanics*. Initially Jammer described von Neumann's attitude on the point in cautious terms:

> As to any details of the process of the first kind, which from now on will be referred to as the 'resolution' or 'reduction' of the superposition ('collapse of the wave packet') or briefly 'reduction', von Neumann was rather *reticent*." (Jammer 1974, 481, emphasis added).

In the very subsequent lines after the above quotation, however, Jammer changes direction. First, he cites the following passage taken from the von Neumann 1932 treatise, concerning the different circumstances in which two distinct modalities of evolution over time for the states of quantum systems apply:

> Now quantum mechanics describes the events which occur in the observed portions of the world, so long as they do not interact with the observing portion, with the aid of the process [of the second kind], but as soon as such an interaction occurs, i.e. a measurement, it requires the application of a process [of the first kind]. (von Neumann 1955, 420)

On the basis of this statement – which does nothing but recall the yet unexplained coexistence of the two dynamical laws in the von Neumann formal framework – and

---

[6] "In plainer language, this assignment entails the following conclusions. If a physical system is in a quantum state which is not an eigenstate of the observable to be measured, then a measurement of that observable causes the system to be suddenly transformed into some eigenstate of the observable. The plausibility of this correspondence between ρ and a measurement is further attested to by the fact that a second measurement following upon the heels of the first can cause no further change in the state of the system, [...] In the sequel I shall speak of the postulate here outlined in connection with the mathematics which suggested it, *as the projection postulate*. It claims that a measurement converts an arbitrary quantum state into an eigenstate of the measured observable." (Margenau 1958, 28-29, emphasis added).

with no further textual support whatsoever Jammer concludes boldly and unhesitatingly:

> This argument for the indispensability of processes of the first kind [i.e. the collapse] also seems to suggest that these processes do not occur in the observed portions of the world, however deeply in the observer's body the boundary is drawn. *They can only occur in his consciousness*. A complete measurement, according to von Neumann's theory, involves therefore the consciousness of the observer. In view of the classification of processes into two mutually irreducible categories, corresponding to the partition of the world into the observed and the observing, which, in spite of the mobility of their dividing line, are also mutually irreducible, *von Neumann's theory is a dualistic one*. (Jammer 1974, 481-482, emphasis added)

Such ambiguities persist in a vast area of the relevant scientific literature. In his 1996 book *Einstein, Bohr and the Quantum Dilemma* Andrew Whitaker presents von Neumann's position on collapse as follows (Whitaker 1996, 199):

> *Where* does collapse take place? Von Neumann gave a clear answer to this question: the final stage of the chain must be entirely different from all previous stages. He said that "it is inherently entirely correct that the measurement or the related process of the subjective perception is a new entity relative to the physical environment and is not reducible to the latter. Indeed, subjective perception leads us into the intellectual inner life of the individual, which is extra-observational by its very nature" [von Neumann 1955, 418]. Von Neumann adds that "experience only makes statements of this type: an observer has made a certain (subjective) observation; and never like this: a physical quantity has a certain value" [von Neumann 1955, 420].

Nothing in the von Neumann text literally allows us to infer that it is the subjective perception that causes the collapse, but Whitaker comments the above quotation by claiming: "So full responsibility for collapse is put onto the observer – *in the capacity of the 'abstract ego'* " (Whitaker 1996, 199), where this 'responsibility' clearly refers to an alleged true causal power of the ego in order for the collapse to actually take place as a real physical process. A very similar ambiguity can be found in a more recent account, due to Franck Laloë, who writes (again, with no uncontroversial textual support) that "von Neumann concludes from his analysis that, indeed, it is not possible to formulate the laws of quantum mechanics in a complete and consistent way without reference to human consciousness." (Laloë 2012, 21, emphasis added).[7]

---

[7] For further textual evidence on the misrepresentation of the von Neumann stance, especially in its alleged role of precursor to London-Bauer about the role of consciousness in the quantum measurement process, see French 2023, 13-16.

In spite of all these unwarranted claims, what von Neumann explicitly *does* say is that "the "how" [of the collapse] remains unexplained for the present" and promises to address this problem: "we shall attempt to bridge this chasm later" (von Neumann 1955, p. 217) [8]. But what does this attempt exactly consist in?

# 3 The von Neumann Model of Quantum Measurement and the Role of the Psycho-Physical Parallelism

The whole Chapter VI of von Neumann's 1932 treatise, the last of the book and entitled *The Measuring Process*, is in fact entirely devoted to this attempt, whose formulation is in some sense 'prepared' by some remarks in the section 1 of the Chapter V. Here von Neumann recalls once again the coexistence of "arbitrary changes" and "automatic changes" (von Neumann 1955, 351), i.e. of the two modalities of evolution in quantum mechanics – (**1**.) denotes the non-unitary evolution, whereas (**2**.) denotes the unitary, Schrödinger evolution – and then writes:

> First of all, it is noteworthy that the time dependence of $H$ is included in **2**., so that one should expect that **2**. would suffice to describe the intervention caused by a measurement: indeed, a physical intervention can be nothing else than the temporary insertion of a certain energy coupling into the observed system, i.e. the introduction of an appropriate time dependency of $H$ (prescribed by the observer). Why then do we need the special process **1**. for the measurement? The reason is this: in the measurement we cannot observe the system $S$ by itself, but must rather investigate the system $S + M$, in order to obtain (numerically) its interaction with the measuring apparatus $M$. *The theory of the measurement is a statement concerning $S + M$, and should describe how the state of S is related to certain properties of the state of M* (namely, the positions of a certain pointer, since the observer reads these). (von Neumann 1955, 352, emphasis added)

On the basis of the emphasized lines, it is quite clear how von Neumann here is endorsing a universality thesis for quantum mechanics, namely (i) the account of the measurement process is an account of the *joint* system $S + M$, (ii) this account is to be given *entirely* in quantum mechanical terms.

---

[8] Among the few who did not incur in the mistake of attributing to von Neumann a causal role for consciousness in quantum measurements we find Olival Freire Junior in his 2015 book *Quantum Dissidents*. In the section 4.2 – entitled *Measurement Problem Before Wigner* – Freire Junior explicitly remarks that von Neumann "refrains from attributing a physical role for the mind in the quantum measurement processes." (Freire Jr. 2015, 144, footnote 6).

Not just that:

> Moreover, *it is rather arbitrary* whether or not one includes the observer in *M*, and replaces the relation between the *S* state and the pointer positions in *M* by the relations of this state and the chemical changes in the observer's eye or even in the brain (i.e. , to that which he has "seen" or "perceived"). We shall investigate this more precisely in VI.1. In any case, therefore, the application of **2**. is of importance only for *S* + *M*. *Of course, we must show that this gives the same result for S as the direct application of* **1**. *on S. If this is successful, then we have achieved a unified way of looking at the physical world on a quantum mechanical basis*. (von Neumann 1955, 352, emphasis added).

Hence, the bulk of the von Neumann treatment of the quantum measurement process in the Chapter VI is exactly to prove that this conjecture holds true, namely that what has been called the 'consistency problem' (Bub 2001, 65) for the two modalities of evolution can be solved.

So, how does von Neumann formulate the consistency problem? Let us divide the world in three parts: the observed system *S*, the measuring apparatus *M*, the observer *O*. Moreover, under the assumption of universality for quantum mechanics, let us suppose two different ways of constructing a joint system out of *S*, *M* and *O*, namely, either the composition <*S* + [*M* + *O*]> or the composition <[*S* + *M*] + *O*>. Now, if **Q** denotes a physical quantity that is subject to measurement, the consistency problem (von Neumann 1955, 420) consists in a question that can be represented as follows:

[*O* + *M*] measures **Q** on *S*          *O* measures **Q** on [*M* + *S*]

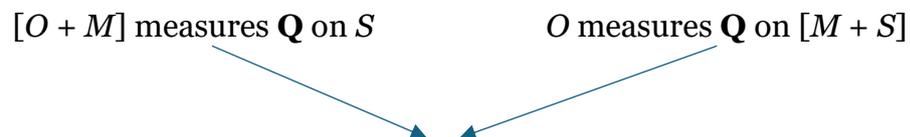

Do these measurements give rise to the *same* **Q**-measurement statistics for *S* ?

Namely: do the performance of a **Q**-measurement on *S* by the system [*O* + *M*] and the performance of a **Q**-measurement on [*M* + *S*] by the system *O* deliver the same **Q**-measurement statistics for *S*? In other words, does the choice on how to compose a joint system of *S*, *M* and *O* affect the **Q**-measurement statistics for *S*? In fact, von Neumann proves that

> applying process 1 directly to [*S*] yields the same density operator for the statistics of all [*S*]-observables after the measurement as we obtain by considering the measurement as a process 2 interaction between [*S*] and [*M*], and then applying process 1 to the measurement of [*S* + *M*] by a second instrument [*O*]." (Bub 2001, 67).

The arbitrariness in the choice of how to decompose the joint system $S + O + M$ had been emphasized in his 1930 book also by Heisenberg:

> There exists a body of exact mathematical laws, but these cannot be interpreted as expressing simple relationships between objects existing in space and time. The observable predictions of this theory can be approximately described in such terms, but not uniquely – the wave and corpuscular pictures both possess the same approximate validity. This indeterminateness of the picture of the process is a direct result of the indeterminateness of the concept 'observation' – it is not possible to decide, *other than arbitrarily*, what objects are to be considered as part of the observed system and what as part of the observer's apparatus. (Heisenberg 1930, 64, emphasis added).

The question, however, was underrated in a Bohrian-sounding fashion:

> Even when this arbitrariness is taken into account the concept 'observation' belongs, strictly speaking, to the class of ideas borrowed from the experiences of everyday life. It can only be carried over to atomic phenomena when due regard is paid to the limitations placed on all space-time descriptions by the uncertainty principle. (Heisenberg 1930, 64).

On the contrary, the solution of this consistency problem in the von Neumann perspective – namely, the mathematical proof of the statistics invariance across possible, different decompositions – has relevant foundational implications. First, it is *this* solution of the consistency problem that leads von Neumann to claim that "we have achieved a unified way of looking at the physical world on a quantum mechanical basis." Second, this solution affects the status of the collapse of the quantum state. In fact, should a difference in the decomposition affect the statistics, the fixation of the boundary between the two systems $S$ and $[M + O]$ in the composition <$S + [M + O]$> or between the two systems $[S + M]$ and $O$ in the composition <$[S + M] + O$> could not be conventional anymore; as a consequence, this would force us to find physically serious reasons to discover *where exactly* should we draw the boundary. Therefore, according to von Neumann, the actual existence of a solution of the consistency problem turns out to be a robust proof that the act of drawing the boundary either between $S$ and $[M + O]$ or between $[S + M] + O$ is a conventional act, devoid of any physical meaning and as such of a totally pragmatic character. But this von Neumann argument on the virtue of the above mentioned arbitrariness is, if not in outright contradiction, at least clearly in tension with the claim that von Neumann had a definite view on the collapse of the wave function as a 'real' physical process: if we take seriously the idea that there is something *real* that *physically* collapses in the measurement interaction, how can the fixation of the boundary either between *S* and [*M* + *O*] or between [*S* + *M*] + *O* be entirely arbitrary?[9]

So we seem to have ample reasons for interpreting the actual von Neumann position as an agnostic one on the status of the collapse itself, with no other qualifications: how then does this agnosticism reflect on the view of a *consciousness-induced* collapse? And what does von Neumann *really* say about consciousness and its role in the quantum measurement process?

To be true, the whole 'theory of consciousness' that can be meaningfully extracted from the 1932 treatise reduces to the pages 418-421– that is, less than one percent of a book that in the English edition counts 445 pages! The already mentioned Chapter VI starts with a short summary of the coexistence of the two laws of evolution of the quantum state – the unitary one and the non-unitary one – after which von Neumann addresses explicitly the relation between measurement and the 'subjective perception':

> It is inherently correct that the measurement or the related process of the subjective perception is a new entity relative to the physical environment and is not reducible to the latter. Indeed, subjective perception leads us into the intellectual inner life of the individual, which is extra-observational by its very nature (since it must be taken for granted by any conceivable observation or experiment). (von Neumann 1955, 418).

This 'extra-observational' subjectivity notwithstanding, it is necessary to provide a correlational formulation of subjective perceptions, so to say, namely to express somehow the content of these perceptions in terms of correlations between outcomes and recording procedures. This is the assumption von Neumann dubs as *principle of psycho-physical parallelism*:

> it is a fundamental requirement of the scientific viewpoint – the so-called principle of psycho-physical parallelism – that it must be possible so to describe the extra-physical process of the subjective perception as if it were in reality in the physical world – i.e., to assign to its very parts equivalent processes in the objective environment, in ordinary space. (von Neumann 1955, 418-419).

After all, it is an entirely reasonable assumption in the context of von Neumann's model: it just requires from quantum theory the possibility to establish a correspondence between a sequence of physical events and a sequence of psychological events, such that the latter constitutes the awareness that a definite measurement occurred, with a definite outcome (Barrett 1999, 47-48). Recalling the above mentioned question of the alternative of where to put the boundary between observed system and observer, either in the composition <*S* + [*M* + *O*]> or in the composition <[*S* + *M*] + *O*>, von Neumann emphasizes again that the location of this boundary "is arbitrary to a large extent" (von Neumann 1955, 420) and interestingly remarks:

> That this boundary can be pushed arbitrarily deeply into the interior of the body of the actual observer is the content of the principle of psycho-physical parallelism – but this does not change the fact that in each method of description the boundary must be put somewhere, if the method is not to proceed vacuously, i.e., if a comparison with experiment is to be possible. *Indeed experience only makes statements of this type: an observer has made a certain (subjective) observation; and never like this: a physical quantity has a certain value.* (von Neumann 1955, 420, emphasis added)

---

[9] For a sustained argument against the view, according to which von Neumann holds that the measurement process produces a physical collapse, see Becker 2004.

This remark turns out to be, in my opinion, a simple re-formulation of the principle of psycho-physical parallelism, that does not ascribe to consciousness any 'causal' power as such in order for the collapse to take place: the arbitrariness on where the boundary between observed system and observer is supposed to be located is taken by von Neumann to support the claim – which is essentially the content of the psycho-physical parallelism – according to which the 'extra-physical process of the subjective perception' and the 'reality in the physical world' should preserve the right kind of match. We can conclude that these passages – that, I repeat, jointly represent the whole 'theory of consciousness' that we can find in the entire von Neuman 1932 treatise – can be hardly taken as supporting the claim that the only way to make sense of the quantum process of measurement (under the universality assumption) is to accept an *active* role of consciousness in driving the evolution from the superposition to a definite result.

## 4  Two Indirect Sources: David Bohm and Eugene Wigner

An interesting, albeit indirect, support to this reading can be found in a source that might appear unexpected, namely 1951 David Bohm's *Quantum Theory*, a book with an interesting story behind (that remains, to a large extent, to be told). At the time of preparation of the book, Bohm was in Princeton, attempting to find his way within a scientific community that – already far from the heated philosophical-sounding discussions of the founding fathers in the first three decades of the century – was not very much concerned anymore with foundational issues surrounding quantum mechanics. *Quantum Theory* was supposed to be an 'orthodox' textbook whose aim, however, was to provide a more readable, robust and consistent version of the theory still within a Copenhagen spirit. As a Bohm biographer writes:

> The Copenhagen explanations were subtle and difficult to follow. Bohm's goal, for his book, was to present them in a language that was simple, clear and concise. It was a Herculean task, since it was not entirely clear that Niels Bohr's arguments themselves were, in fact, totally consistent – at some deeper level there might be confusions and inconsistencies. In any event, with the help of the critical reactions of his friends, Bohm attempted to present Bohr in the best possible light. (Peat 1997, 87).[10]

The effort to minimize the ambiguities of the Copenhagen attitude, jointly with an unusual sentitivity for the philosophical dimension of physics, make of *Quantum Theory* a book worth investigating in its own right[11]: curiously enough, the most sophisticated version of the Copenhagen quantum mechanics was bound to be proposed by the future most fierce enemy of the quantum orthodoxy!

For the sake of the present paper, our focus will be on the Bohm treatment of the measurement process in quantum mechanics, developed in the Chapter 22. This treatment, although presented as differing from the von Neumann one, will be shown to depend on several presuppositions of the latter: in unfolding these presuppositions at the service of his own formulation, the Bohm treatment in fact turns out to reinforce the claim according to which von Neumann was far from subscribing to a view of the quantum measurement process that relies on the presence of conscious observers in order to be consistent.

The title of the Chapter 22 of Bohm's book – *Quantum Theory of the Measurement Process* – clearly announces the underlying 'universalistic' attitude, expressed in the very first lines:

> *If the quantum theory is to be able to provide a complete description of everything that can happen in the world, however, it should also be able to describe the process of observation itself in terms of the wave functions of the observing apparatus and those of the system under observation.* Furthermore, in principle, it ought to be able to describe the human investigator as he looks at the observing apparatus and learns what the results of the experiment are, this time in terms of the wave functions of the various atoms that make up the investigator, as well as those of the observing apparatus and the system under observation. *In other words, the quantum theory could not be regarded as a complete logical system unless it contained within it a prescription in principle for how all of these problems were to be dealt with*. (Bohm 1951, 583, emphasis added).

---

[10] The book received a favorable attention – Pauli praised the integration of physics, mathematics and philosophy exhibited by Bohm's treatment (Peat 1997, 109) – but Bohr remained silent about it.
[11] The peculiar and sophisticated way in which Bohm undertook the task of elaborating an 'orthodox' presentation of quantum theory would deserve a more detailed analysis: few scattered remarks on this point can be found in Freire Jr. 2015, 47-51.

Immediately after, warning that "in this chapter, we shall show how one can treat these problems within the framework of quantum theory", Bohm adds a footnote where he refers to 1932 von Neumann's treatise as a text that proposes "*another* treatment of this problem" (Bohm 1951, 583, emphasis added) and, in fact, this is the only occurrence of von Neumann's name in the whole book: but what is it exactly that distinguishes Bohm's approach from the von Neumann one?

The above assumption – according to which "quantum theory must be able to describe the process of observation itself" – implies addressing somehow the crucial event in a quantum measurement: namely, the transition from the entangled state of the pair <Observed System $S$ + Measuring Apparatus $A$> to the (non-entangled) state corresponding to the recording of a result of the measurement. Unlike von Neumann, who introduced explicitly in his account a duality of dynamical evolutions, corresponding to correlation without measurement (unitary, Schrödinger evolution) on one side and to correlation with measurement (non-unitary evolution) on the other, Bohm appears to rely on a Schrödinger kind of evolution law, a circumstance that leads to what is known as the measurement problem in quantum mechanics: how does Bohm circumvent this implication? By assuming that, *in practice*, "all real observations are, in their last stages, classically *describable*" (Bohm 1951, 585), an option that shows effectively why after all the Bohm treatment in the book is a *Copenhagen-style* exercise in interpretation[12]. Namely, the universality of quantum theory requires to describe the <$S + A$> interaction as a quantum process but the output of the interaction – which is meant to be a *unique result* – must receive a classical description:

> [...] at the quantum level of accuracy the entire universe (including, of course, all observers of it), must be regarded as forming a smgle indivisible unit with every object linked to its surroundings by indivisible and incompletely controllable quanta. If it were necessary to give all parts of the world a completely quantum mechanical description, a person trying to apply quantum theory to the process of observation would be faced with an insoluble paradox. This would be so because he would then have to regard himself as something connected inseparably with the rest of the world. On the other hand, the very idea of making an observation implies that what is observed is totally distinct from the person observing it. This paradox is avoided by taking note of the fact that all real observations are, in their last stages, classically describable. The observer can therefore ignore the indivisible quantum links between himself and the classically describable part of the observing apparatus from which he obtains his information, because these links produce effects that are too small to alter in any essential way the significance of what he sees. (Bohm 1951, 584-585).

---

[12] In the text Bohm appeals to the notion of *random phase factor*, referred to the coupling process involving both the state of the observed system and the state of the measuring apparatus (cp. Freire 2019, 50).

It should be emphasized that this 'classical describability' has a pragmatic, epistemological and *non-ontological* character, under the universalistic hypothesis that it is quantum theory that in principle covers every natural phenomenon, observation included. In this respect, very consequentially Bohm entitles the immediately subsequent section *Extent of Arbitrariness in Distinction Between the Observer and What He Sees* (Bohm 1951, 586), a section which recalls in nearly literal terms the part of the 1932 treatise in which von Neumann conceives 'arbitrariness' in terms of the equivalence between <*S* + [*M* + *O*]> and <[*S* + *M*] + *O*>:

In the event that there is more than one classically describable stage of the apparatus, any one stage may be chosen as the point of separation between the observer and what he sees. Consider, for example, an experiment in which a person obtains information for a photograph. One possible description of this experiment is as follows: The system observed consists of the objects photographed, plus the camera, plus the light that connected object with image. The observer is then said to obtain his information by looking at the plate. Because this process is classically describable, there is a sharp distinction between the observer and the plate that he is looking at. An equally good description, however, involves regarding the system under investigation as the object itself. The camera and the plate can then be considered as part of the observer. A third description of the same procedure would be to say that the investigator observes the image on the retina of his eye, so that the retina of the eye, plus the rest of the world, including of course the photographic plate, are to be regarded as the system under observation. (Bohm 1951, 586).

On the basis of this claim of arbitrariness, Bohm is led to a conclusion which is very similar in spirit to the conclusion to which von Neumann is led by the solution to what he had taken to be a consistency problem, namely the invariance of the statistics across different choices in decomposing the system <*System + Measuring Apparatus + Observer*>:

[about] the problem of how far into the brain the point of distinction between the observer and what is observed can be pushed […] *we wish to stress that the question is completely irrelevant as far as the theory of measurements is concerned since, as we*

*have already seen, it is necessary only to carry the analysis to some classically describable stage of the apparatus.*" (Bohm 1951, 587).

Moreover, he closes the section with a remark that stresses the irrelevance of the details in the neurobiological structure of the observer, in order for the consistency of the quantum theory desciption of the observation process:

> In view of the fact that we know so little about the details of the functioning of the brain, it seems fortunate that the analysis has to be carried only to a classically describable part of the apparatus. (Bohm 1951, 588).

If our working hypotesis is correct – namely, that Bohm in his role of Copenhagen interpreter actually takes from von Neumann the main ingredients in the description of the relations among the measured system, the measuring apparatus and the observer, in spite of the lack of distinction between the two different modalities of evolution – then it may argued that *at least at the time of the book preparation by Bohm* the von Neumann view of the measurement process in quantum mechanics was *not* widely taken as a view that included consciousness as a causal factor (whatever *causal* might reasonably mean) in the process that at the end of a measurement interaction produces a result on the apparatus side.

And what about the *locus classicus* on the apparently unavoidable role of consciousness for the consistency of the account of the measurement process in quantum mechanics, namely, the Eugene P. Wigner article "Remarks on the Mind-Body Question" (1961)? Here Wigner overtly claims that the theoretical role of consciousness emerges in the very circumstance of classical physics being supplanted by quantum physics:

> When the province of physical theory was extended to encompass microscopic phenomena, through the creation of quantum mechanics, the concept of consciousness came to the fore again: it was not possible to formulate the laws of quantum mechanics in a fully consistent way without reference to the consciousness. All that quantum mechanics purports to provide are probability connections between subsequent impressions (also called "apperceptions") of the consciousness […] (Wigner 1961, 172).

In order to support his main claim Wigner mentions for instance a statement that Werner Heisenberg made in an article published in 1958 – "The laws of nature which we formulate mathematically in quantum theory deal no longer with the particles themselves but with our knowledge of the elementary particles" (Heisenberg 1958, 99)

– referring to which Wigner clearly equates *knowledge* with *consciousness*, and goes very far in this path by characterizing the task itself of physics as follows:

> One realises that *all* the information which the laws of physics provide consists of probability connections between subsequent impressions that a system makes on one if one interacts with it repeatedly, i.e., if one makes repeated measurements on it. The wave function is a convenient summary of that part of the past impressions which remains relevant for the probabilities of receiving the different possible impressions when interacting with the system at later times. (Wigner 1961, 174, emphasis in the original).[13]

Yet, it is important to emphasize that nowhere Wigner enlists von Neumann as a forerunner of the thesis of the necessity of consciousness in order for the quantum theory of measurement to be consistent. All Wigner says in this respect is that the main task of quantum mechanics is providing

> probability connections between subsequent impressions (also called "apperceptions") of the consciousness, *and even though the dividing line between the observer, whose consciousness is being affected, and the observed physical object can be shifted towards the one or the other to a considerable degree*, it cannot be eliminated. (Wigner 1961, 172, emphasis added).

A footnote is appended to the emphasized lines with a reference to the Chapter VI of von Neumann's 1932 treatise, where – as we have seen in our section 3 above – he exploits the arbitrariness on where to locate the boundary between the measured system, the measuring apparatus and the observer in order to neutralize the lack of explanation of the collapse of the quantum state in the event of a measurement.[14]

## 5   Conclusions

In the above pages I have attempted to substantiate what appears to be a rather cautious attitude by von Neumann on such a deep interpretational question as the need of the notion of observer in order for the quantum description of measurement to be consistent. The von Neumann attitude, however, has been long misrepresented by the large majority of the subsequent literature, that turned out to credit him with a much more radical view: according to it, not only the collapse was in itself a truly physical process, but also the only way to accomodate it within a quantum description of a typical measurement was the introduction of human consciousness as a kind of 'causal' factor. On the basis of the work of reconstruction pursued by the phenomenological reading of the London-Bauer approach, started by Steven French more than twenty years ago, and on the basis of analyses like the one I have proposed in the present work, we may think that the time is ripe to tell a more balanced story on the relation between the notion of consciousness and the foundations of quantum mechanics in the work of the first scientist – János von Neumann – who explicitly and rigorously addressed the implication of a really *universal* formulation of quantum physics.

---

[13] As is now well-known, this Wigner approach was one of the main sources of the Qbist interpretation of quantum mechanics (Caves, Fuchs, Schack 2007, Fuchs, Mermin, Schack, R. 2014).

[14] The Wigner article contains also a quick reference to the emphasis that London and Bauer place on the role of consciousness in their 1939 work (Wigner 1961, 173, fn 5).